\documentclass[lettersize,journal]{IEEEtran}
\usepackage{graphicx}
\usepackage{latexsym}
\usepackage{amsmath}
\usepackage{amsthm}
\usepackage{makecell,rotating,multirow,diagbox}
\usepackage{amsfonts}
\usepackage{subfigure}
\usepackage{dsfont}
\usepackage{epstopdf}
\usepackage{threeparttable}
\usepackage{multirow}

\usepackage{enumerate}

\usepackage{epsfig}

\usepackage{bm}

\usepackage{cite}

\usepackage{algorithmic}

\usepackage{textcomp}
\usepackage{xcolor}
\usepackage{algorithm}
\usepackage{algorithmic}
\usepackage{setspace}

\usepackage{amssymb}
\usepackage{amsmath}
\usepackage{tabu}
\usepackage{makecell}

\ifCLASSINFOpdf
\else
\fi
\begin{document}

\title{{Superimposed Pilot-based Channel Estimation for RIS-Assisted IoT Systems Using Lightweight Networks} }
\author{{Chaojin~Qing,~\IEEEmembership{Member,~IEEE,}
       Li~Wang,
       Lei~Dong,
       Guowei Ling,
       and~Jiafan~Wang}

\thanks{C. Qing, L. Wang, L. Dong, G. Ling and {J. Wang} are with the School of Electrical Engineering and Electronic Information, Xihua University, Chengdu, 610039, China (E-mail: qingchj@mail.xhu.edu.cn). }
}


\maketitle

\begin{abstract}
Conventional channel estimation (CE) for Internet of Things (IoT) systems encounters challenges such as low spectral efficiency, high energy consumption, and blocked propagation paths. Although superimposed pilot-based CE schemes and the reconfigurable intelligent surface (RIS) could partially tackle these challenges, limited researches have been done for a systematic solution. In this paper, a superimposed pilot-based CE with the reconfigurable intelligent surface (RIS)-assisted mode is proposed and further enhanced the performance by networks. Specifically, at the user equipment (UE), the pilot for CE is superimposed on the uplink user data to improve the spectral efficiency and energy consumption for IoT systems, and two lightweight networks at the base station (BS) alleviate the computational complexity and processing delay for the CE and symbol detection (SD). These dedicated networks are developed in a cooperation manner. That is, the conventional methods are employed to perform initial feature extraction, and the developed neural networks (NNs) are oriented to learn along with the extracted features. With the assistance of the extracted initial feature, the number of training data for network training is reduced. Simulation results show that, the computational complexity and processing delay are decreased without sacrificing the accuracy of CE and SD, and the normalized mean square error (NMSE) and bit error rate (BER) performance at the BS are improved against the parameter variance.

\end{abstract}


\section{Introduction}
\IEEEPARstart{A}{s} the cornerstone of the future Internet of Things (IoT) connectivity, the evolution of fifth-generation (5G) and sixth-generation (6G) networks has attracted consistent attention in the application of IoT. For example, intelligent buildings connected with the internet to manage different devices \cite{r34}, smart health care and intelligent driving proposed by \cite{r33}, and home automation put forward by \cite{r35}, etc. In these IoT systems, channel estimation (CE) plays critical roles, such as to overcome channel time variation \cite{r37} or the increase of occlusion probability, and adjust to an affordable transmission power using appropriate modulation and coding methods \cite{r36}.

The CE for IoT systems is vital for effective receiver operation \cite{r37}{, \cite{r39} and \cite{r38}}. In IoT downlink systems, a pilot-based hybrid CE method is introduced in \cite{r37}, which  combines the 1-D time-domain Wiener filter technique with a computationally simple maximum likelihood estimator in the frequency domain, and an improved computationally efficient linear minimum mean square error (MMSE) estimator for the downlink IoT systems is proposed in \cite{r38}. As for IoT uplink systems, the least-squares (LS) and MMSE based CE is adopted in \cite{r39}. \cite{r50} proposes a low complexity CE algorithm based on the conventional LS method for downlink narrow-band IoT systems. \cite{r51} discusses the downlink CE of broadband IoT systems. Given these CE schemes in \cite{r37} and \cite{r39,r38,r50,r51}, there is still room for improving the spectral efficiency and energy consumption. On the one hand, existing pilot-based CEs for an IoT system must allocate additional spectrum resources to transmit pilots, and this causes low spectral efficiency. On the other hand, energy-consuming needs to be handled in IoT systems. One extreme case in \cite{r40} is that the user equipment (UE) aims to extend up to ten years' battery lifetime. In this situation, transmitted pilots and data of the IoT system separately will definitely increase the energy consumption and hardly achieve the system target. Without extra time-frequency resources for the pilot, the strategy in \cite{r41} transmits the pilot and data in a superimposed manner, which alleviates the issue of low spectral efficiency and high energy consumption, and this inspires us to propose a CE solution of IoT systems based on superimposed pilots.


Besides, it is a common situation that IoT communication is blocked due to complex scenarios of propagation paths in industrial IoT (IIoT) \cite{r37}. Thus, increasing the robustness of the communication link is an urgent task to guarantee CE performance. To resolve blocked propagation paths, reconfigurable intelligent surface (RIS) provides an attractive option \cite{r2}. The RIS, an artificial panel of electromagnetic material, is made from a large array of low-cost passive scattering elements, which can manipulate the wireless environment by adjusting the amplitude or phase shift of reflected signals \cite{r2}. Different from the traditional amplify-and-forward relay, the passive elements in RIS consume little energy \cite{r3}. {From \cite{r49}, the influence of material used in the RIS is another interesting topic, {which gives us a novel perspective for the channel estimation in future works. By considering a passive RIS, we focus on the superimposed pilot-based channel estimation for RIS-assisted communication systems in this paper.} In recent years, embedding RIS in the IoT systems has been envisioned as a revolutionary means to transform a passive wireless communication environment into an active reconfigurable one, which can provide environmental intelligence for different communication objectives \cite{r4}. In addition, RIS also enhances system throughput by at least 40 percent \cite{r5} and system coverage by 1/3 \cite{r6}. Then, deploying RIS for the superimposed pilot-based CE of IoT systems is a highly desired approach to tackle the issue of blocked propagation paths, which, however, has not been well investigated in existing works.

In recent years, deep learning (DL) has made a major breakthrough in advanced information processing and computer vision\cite{r45}. In \cite{r44}, the essence of DL is to learn the mapping relationship between input and output through training data samples, get a model structure, and then feed the test data to obtain the predicted output via the model. Even so, it is still hard to directly explain the internal mechanism and theoretical analysis of DL\cite{r45}, \cite{r44}. Potential applications of DL in the physical layer have been increasingly recognized due to the new features of future communications, such as complex scenarios of unknown channel models and precise processing requirements\cite{r45}. In addition, DL-based CE in RIS-assisted communication systems has also aroused extensive research interest. \cite{r42} proposed two convolutional neural networks (CNN)-based methods to execute the de-noising and approximate the optimal MMSE {CE} solution.
 \cite{r17} proposed an enhanced extreme learning machine (ELM)-based {CE} to facilitate accurate {CE}.
 \cite{r43} proposed an untrained deep neural network (DNN) based on the deep image prior network to de-noise the effective channel of the system acquired by the conventional pilot-based LS estimation and obtain a more accurate estimation. However, DL-based superimposed CE in RIS-assisted communication systems has not been investigated{, which is particularly important for an IoT system to reduce the energy consumption with high spectral efficiency.}

To reduce the energy consumption while maintaining the spectral efficiency of IoT systems, tackling blocked propagation paths, and enhancing the CE's accuracy, we investigate the superimposed pilot-based CE for RIS-assisted {IoT systems} in this paper.
The main contributions of our work are summarized as follows:

\begin{enumerate}
  \item We propose a superimposed pilot-based and RIS-assisted mode into the IoT system to alleviate the issues of spectral
efficiency and energy-consuming. Besides, the non-superimposed pilot-based CE in \cite{r38} encounters the issue that its pilot/data cannot be received completely, and the superimposed pilot-based method can effectively alleviate this {issue (especially when the channel changes frequently)}. On the one hand, by employing the superimposed pilot-based mode, we reduce the energy consumption and improve the spectral efficiency of IoT systems. On the other hand, the robustness of the communication link is enhanced by employing RIS. Especially, the combinations of superimposed pilot-based mode and RIS further reduce energy consumption, and thus prolong the battery life of UE. As far as we know, with prolonged battery life of UE and enhanced spectral efficiency of the IoT system, the issue of improving the accuracy of the CE at the BS has not been well addressed in \cite{r40,r48}.
      Therefore, it is beneficial to study the superimposed pilot-based CE for RIS-assisted systems.

  \item We develop two dedicated lightweight networks to reduce the computational complexity and processing delay for the CE and symbol detection (SD) at the BS. From the perspective of integrating the non-NN and NN-based solutions, the initial features are highlighted by employing conventional estimation and detection methods to perform feature extraction, and the lightweight networks are oriented to learn along with the highlighted initial features. Thus, the non-NN and NN-based solutions cooperatively improve the CE and SD, while holding the lightweight for the developed networks. Due to the assistance of non-NN mode, both the CE and SD networks are shallow networks and thus have lightweight structure. The computational complexity of proposed method is lower than the conventional method, e.g., MMSE channel estimation and MMSE equalization, which saves the computational resources and shortens the processing delay of BS.

  \item With the reduced computational complexity and processing delay by using de-noising network, feature extraction, and feature fusion, we further improve the NMSE and BER performances at the BS. For CE, we exploit the learning ability of developed CE network according to de-noising (suppress the superimposed interference and noise) and feature extraction, which alleviates the influence of superimposed interference. The improved CE refines NMSE performance, and thus improves its subsequent SD. Besides, the developed fusion network captures the additional features for SD and improves the BER performance effectively at the BS.

\end{enumerate}

The remainder of this paper is structured as follows: In Section II, we present the system model of superimposed pilot-based CE. The proposed method is presented in Section III. The computational complexity is analyze in Section IV and followed by numerical results in Section V. Finally, Section VI concludes our work.

\textit{Notations}: Bold face lower case and upper case letters represent vector and matrix, respectively. ${\left(\cdot \right)^\mathrm{T}}$ is the transpose. $ \odot $ stands for the Hadamard product. ${\mathop{\mathrm{Re}(\cdot)}} $ and ${\mathop{\mathrm{Im}(\cdot)}}$ represent the real and imaginary parts of complex numbers, respectively.

\begin{figure}[t]
\centering
\includegraphics[scale=0.3]{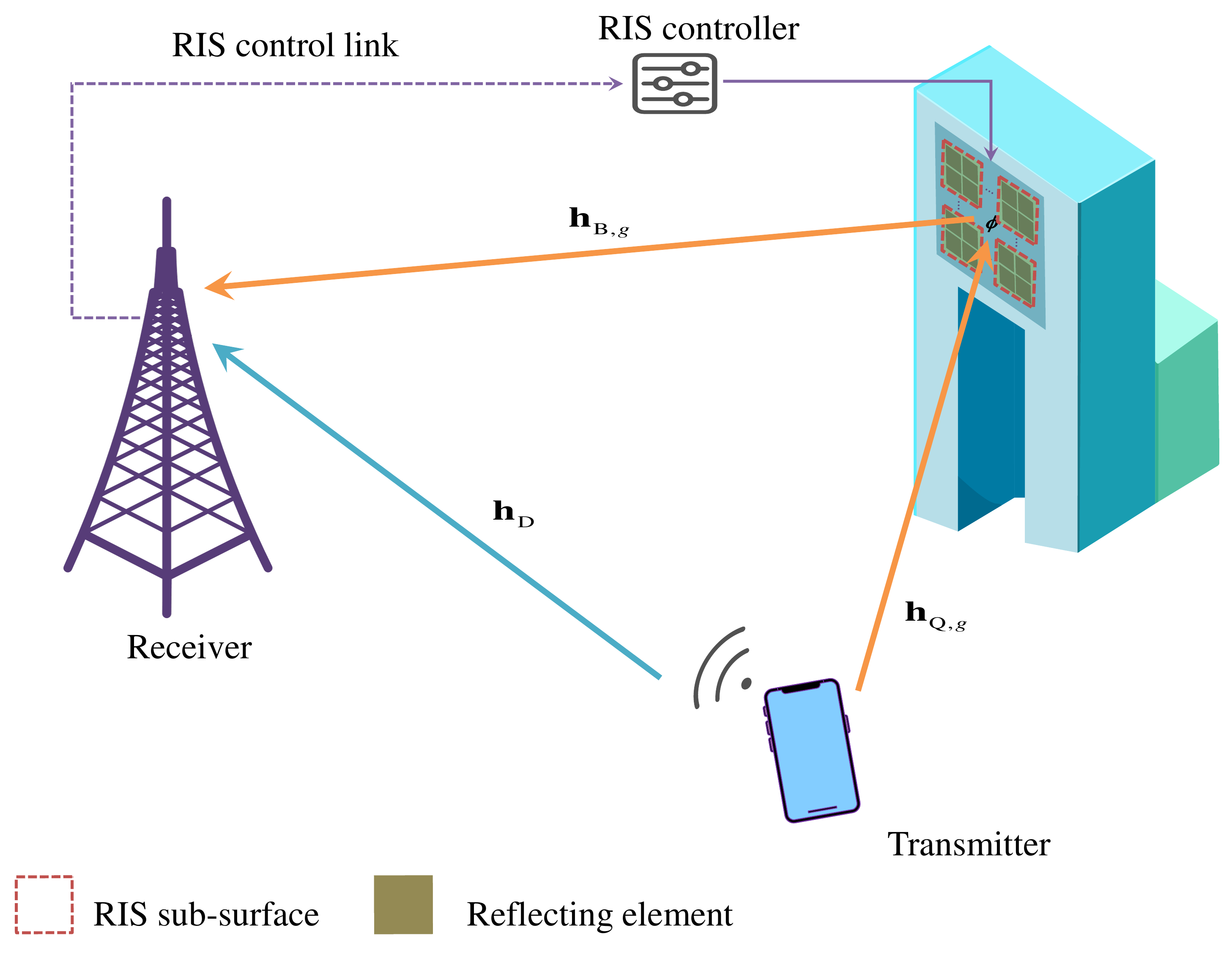}
\caption{An illustration of RIS-assisted OFDM communication in the uplink.}
\label{figure0}
\end{figure}

\begin{figure*}[t]
\centering
\includegraphics[scale=0.66]{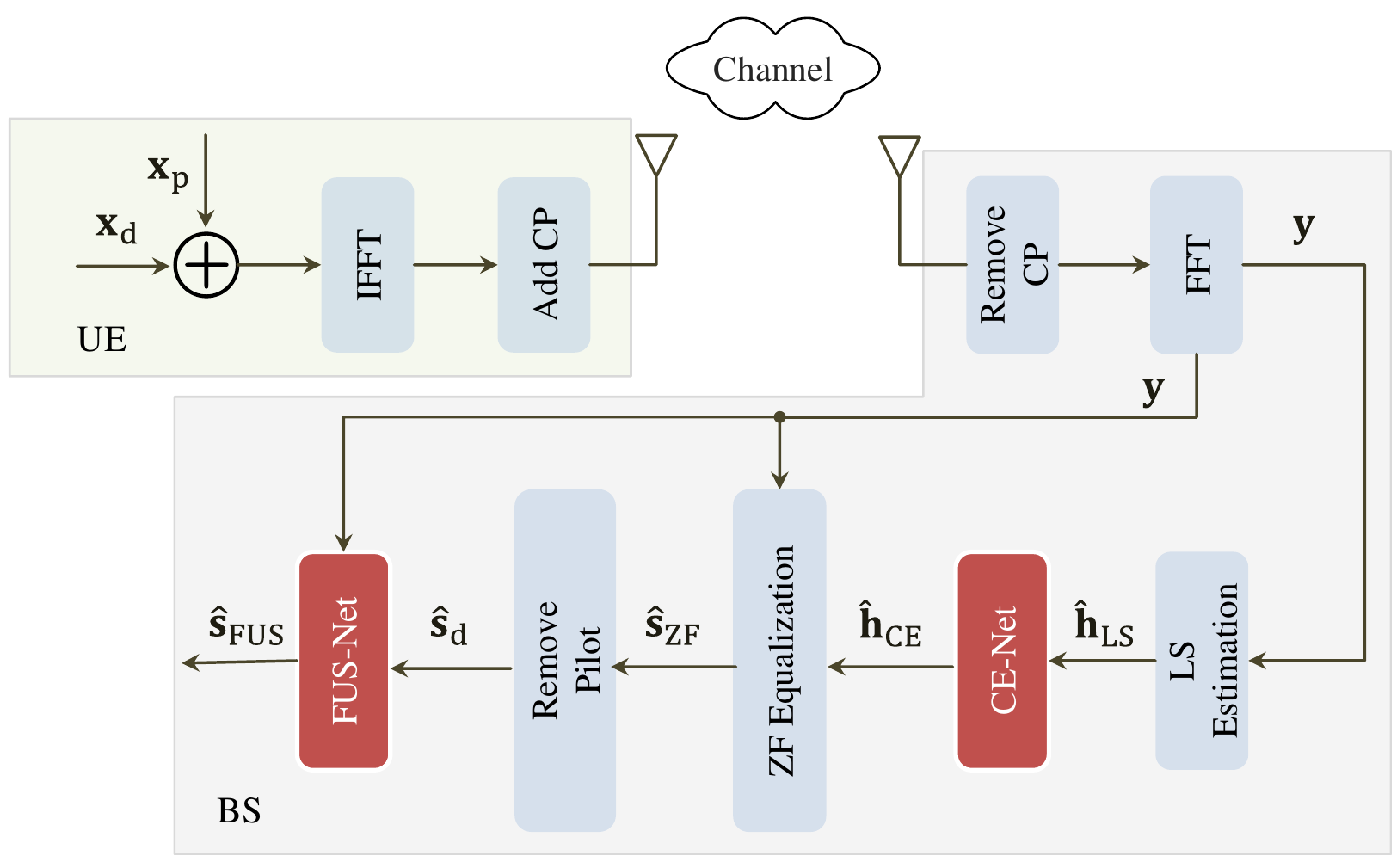}
\caption{Architecture of Proposed Method.}
\label{figure2}
\end{figure*}
\section{System Model}

 As shown in {Fig.~\ref{figure0}}, we consider a frequency-selective Rician fading RIS-assisted {IoT system with OFDM modulation}. In {Fig.~\ref{figure0}}, supposing the propagation path is blocked by buildings, the RIS is installed on the surface of the building to alleviate this issue. $\mathbf{h}_{\mathrm{D}}$ denotes the composite channel frequency response (CFR) of transmitter-receiver link.
 $\mathbf{h}_{\mathrm{B},g}$ and $\mathbf{h}_{\mathrm{Q},g}$ represent the aggregated CFRs of the RIS-receiver link and transmitter-RIS link related to the $g$-th sub-surface, respectively. The RIS is composed of many passive reflecting elements, and to reduce the complexity and training overhead of {CE}, adjacent elements are grouped into a sub-surface to share a common reflection coefficient\cite{r14}. Besides, the RIS control link is used to adjust the phase shift.
 This system considers $N$ sub-carriers and assumes the maximum delay spread $L$ is shorter than the cyclic prefix (CP) length $L_{\mathrm{CP}}$, i.e., $L < L_{\mathrm{CP}}$\cite{r39}, \cite{r14}, to resist inter symbol interference (ISI) and inter carrier interference (ICI). The frequency-domain signal received at the receiver is expressed as
\begin{equation}\label{EQ1}
{\bf{y}} = \sqrt {\lambda P} {\bf{h}} \odot {{\bf{x}}_{\rm{p}}} + \sqrt {\left( {1 - \lambda } \right)P} {\bf{h}} \odot {{\bf{x}}_{\rm{d}}} + {\bf{w}},
\end{equation}
where $\lambda  \in \left[ {0,1} \right]$ is the power proportional coefficient, $P$ stands for the total transmitting power. $\mathbf{h} = {\left[ {{h_1},{h_2}, \cdots ,{h_N}} \right]^T}$ represents the CFR between the receiver and transmitter. $\mathbf{x}_\mathrm{p}\in \mathbb{C}^{N \times 1}$ is the pilot and ${\mathbf{x}_\mathrm{d}} \in \mathbb{C}^{N \times 1}$ denotes the modulated signal. $\mathbf{w} \in \mathbb{C}^{N \times 1}$ is the circularly symmetric complex Gaussian (CSCG) distribution with mean zero and variance $\sigma _w^2$.

The composite CFR between the receiver and transmitter is given as\cite{r14}
\begin{equation}\label{EQ2}
\mathbf{h} = {\mathbf{h}_{\mathrm{D}}} + {\mathbf{H}_{\mathrm{QB}}}{\bm{\phi}} ,
\end{equation}
where ${\mathbf{h}_{\mathrm{D}}} \in \mathbb{C}^{N \times 1}$ denotes the CFR of the transmitter-receiver link, and ${\mathbf{H}_{\mathrm{QB}}} \in \mathbb{C}^{N \times M}$ represents the equivalent cascaded CFR of the transmitter-RIS-receiver link. $\bm{\phi}  \buildrel  \over = \left[ {{\phi _1},{\phi _{2,}} \cdots ,{\phi _G}} \right]^T$ stands for the phase-shift vector, which is given by
\begin{equation}\label{EQ3}
{\phi _g} = {\alpha _g}{e^{j{\theta _g}}},\quad g = 1, \cdots ,G,
\end{equation}
where ${\theta _g} \in \left[ {0,2\pi } \right]$ denotes the phase shift of the $g$-th sub-surface, $G$ is the number of sub-surface. To simplify the design of hardware and maximize the reflection power of the RIS, we fix ${\alpha _g} = 1,\forall g = 1, \cdots ,G$ and only adjust the phase shift ${\theta _g}$\cite{r17}.

By denoting ${\mathbf{H}_{\mathrm{QB}}} = \left[ {{\mathbf{h}_{\mathrm{QB},1}},{\mathbf{h}_{\mathrm{QB},2}}, \cdots ,{\mathbf{h}_{\mathrm{QB},G}}} \right]$, ${\mathbf{h}_{\mathrm{QB},g}}$ {is} expressed as
\begin{equation}\label{EQ4}
{\mathbf{h}_{\mathrm{QB},g}} = {\mathbf{h}_{\mathrm{Q},g}} \odot {\mathbf{h}_{\mathrm{B},g}},
\end{equation}
where ${{\bf{h}}_{\mathrm{Q},g}} \in \mathbb{C}^{N \times 1}$ and ${{\bf{h}}_{\mathrm{B},g}} \in \mathbb{C}^{N \times 1}$ represent the aggregated CFRs of the transmitter-RIS link and RIS-receiver link related to the $g$-th sub-surface, respectively.

According to (\ref{EQ1}), (\ref{EQ2}) and (\ref{EQ4}), the received signal in the frequency domain is rewritten as
\begin{equation}\label{EQ5}
\begin{array}{l}
\mathbf{y} = \left( {{{\bf{h}}_{{\rm{D}}}} + \sum\limits_{g = 1}^G {{{\bf{h}}_{{\rm{Q}},g}}{\phi _g} \odot {{\bf{h}}_{{\rm{B}},g}}} } \right)\odot\\
\quad\quad \,\left( {\sqrt {\lambda P} {\mathbf{x}_\mathrm{p}} + \sqrt {\left( {1 - \lambda } \right)P} {\mathbf{x}_\mathrm{d}}} \right) + \mathbf{w}
\end{array}.
\end{equation}
With the received signal $\mathbf{y}$, the LS estimation and ZF equalization are used to highlight the {initial features} of estimation and alleviate the network learning, respectively.

In this paper, to save bandwidth resources {and energy-consuming}\cite{r8}, we adopt the method of superimposed pilot for {CE} and SD in Fig.~\ref{figure2}. {Two dedicated lightweight networks, namely CE-Net and FUS-Net, are developed to implement CE and SD}, respectively. Distinguished from the conventional {methods, e.g., the MMSE CE and MMSE SD,} non-NN and NN-based approaches are {integrated} into our work, in which the CE-Net and FUS-Net are embedded into {these conventional methods} to {cooperatively} improve the performance of CE and SD.

\section{{Superimposed Pilot-based Channel Estimation}}
As shown in Fig.~\ref{figure2}, first, we superimpose the pilot $\mathbf{x}_\mathrm{p}$ and modulated signal $\mathbf{x}_\mathrm{d}$ together at the UE. Second,
we perform inverse fast Fourier transform (IFFT) and add CP operations. Third, the signal propagates over the wireless channel. At the BS, {the received signal $\mathbf{y}$ is achieved after removing CP and performing fast Fourier transform (FFT) operations.}
{Next}, the conventional LS estimation is employed to highlight the initial features of CE for the lightweight NN. Since a lightweight NN possesses very limited learning ability, the highlighted initial features orient the learning of CE-Net and thus improve the effectiveness of CE.
{Similarly, the developed FUS-Net is also a lightweight network and thus needs to extract the initial equalization features. In this paper, the conventional ZF equalization is employed as a feature extractor to capture the initial equalization feature $\widehat{\mathbf{s}}_{\mathrm{ZF}}$. With the initial equalization feature $\widehat{\mathbf{s}}_{\mathrm{ZF}}$, the coarse data $\widehat{\mathbf{s}}_\mathrm{d}$ is obtained by cancelling the superimposed pilot.~Then, the coarse data $\widehat{\mathbf{s}}_\mathrm{d}$ and the received signal $\mathbf{y}$ are fed into} the FUS-Net to produce the detected symbol $\widetilde{\mathbf{s}}_{\mathrm{FUS}}$. {In Section III-A, the initial feature extraction for CE-Net is presented. {Then}, we develop a lightweight NN, named as CE-Net, to improve the performance of CE in Section III-B.} In Section III-C, the initial feature extraction for FUS-Net is elaborated. Next, fusion learning-based lightweight NN, named as FUS-Net, is used to refine the performance of SD in Section III-D. Last, in Section III-E, the details of online deployment are described.

\begin{algorithm}[t]
\begin{spacing}{0.8}
\caption{Fusion learning{-based CE and SD}}
\begin{spacing}{1.2}
\end{spacing}
\hspace*{0.02in}{\bf Input:}
Initial estimation $\mathbf{\widetilde{h}}_{\mathrm{LS}}$, training learning rate of \\
\hspace*{0.45in}CE-Net: $\gamma_1$, training learning rate of FUS-Net: $\gamma_2$,
\hspace*{0.4in} batch size: $\nu$, number of gradsteps for CE-Net: $G_{\mathrm{CE}}$,
\hspace*{0.4in} number of gradsteps for FUS-Net: $G_{\mathrm{FUS}}$.
\begin{spacing}{1.1}
\end{spacing}
\hspace*{0.02in}{\bf Output:}
Refined detection $\mathbf{\widetilde{s}}_{\mathrm{FUS}}$.

\begin{spacing}{1.2}
\end{spacing}
\begin{spacing}{0.8}
\end{spacing}
\hspace*{0.03in}{\bf Training phase:}
\begin{spacing}{-0.9}
\end{spacing}
\begin{algorithmic}[1]

\begin{spacing}{2.0}
\end{spacing}
\STATE {Randomly initialize the network parameters $\mathrm{\Theta}_{\mathrm{CE}}$  and $\mathrm{\Theta}_{\mathrm{FUS}}$.}
\STATE {Generate the training set $\left\{ {{\mathbf{\widetilde{h}}_{\mathrm{LS}}},{\mathbf{\widetilde{h}}_{\mathrm{Label}}}} \right\}$ and $\left\{ {{\mathbf{\widetilde{s}}_\mathrm{in}},{\mathbf{\widetilde{x}}_\mathrm{d}}} \right\}$.}
\par \setlength\parindent{1em}\FOR{$t=1, ..., G_{\mathrm{CE}}$}
\STATE {Randomly select $\nu$ training samples from $\left\{ {{\mathbf{\widetilde{h}}_{\mathrm{LS}}},{\mathbf{\widetilde{h}}_{\mathrm{Label}}}} \right\}$ as the training batch.}
\begin{spacing}{0.5}
\end{spacing}
\STATE {Update $\mathrm{\Theta}_{\mathrm{CE}}$ by using the Adam algorithm (learning rate $\gamma_1$) to minimize $Loss_{\mathrm{CE-Net}}$}.

\ENDFOR\\

\par \setlength\parindent{1em}\FOR{$t=1, ..., G_{\mathrm{FUS}}$}
\STATE {Randomly select $\nu$ training samples from $\left\{ {{\mathbf{\widetilde{s}}_\mathrm{in}},{\mathbf{\widetilde{x}}_\mathrm{d}}} \right\}$ as the training batch.}
\STATE {Update $\mathrm{\Theta}_{\mathrm{FUS}}$ by using the Adam algorithm (learning rate $\gamma_2$) to minimize $Loss_{\mathrm{FUS-Net}}$}.

\ENDFOR\\

\begin{spacing}{1.3}
\end{spacing}
\hspace*{-0.34in}{\bf Testing phase:}
\begin{spacing}{0.4}
\end{spacing}
\STATE {Load the trained parameters $\mathrm{\Theta}_{\mathrm{CE}}$ and $\mathrm{\Theta}_{\mathrm{FUS}}$.}

\STATE{Perform LS estimation to obtain $\mathbf{\widehat{h}}_{\mathrm{LS}}$ using Eq.~(\ref{EQ6}).}
\STATE {Reshape the complex-valued $\mathbf{\widehat{h}}_{\mathrm{LS}}$ to real-valued $\widetilde{\mathbf{h}}_{\mathrm{LS}}$ using Eq.~(\ref{EQ7}).}
\STATE {Predict  $\mathbf{\widetilde{h}}_{\mathrm{CE}}$ based on $\mathrm{\Theta}_{\mathrm{CE}}$ and $\mathbf{\widehat{h}}_{\mathrm{LS}}$ using Eq.~(\ref{EQ8}).}
\STATE {Perform ZF equalization to obtain $\mathbf{\widehat{s}}_{\mathrm{ZF}}$ using Eq.~(\ref{EQ10}).}
\STATE {Cancel the superimposed interference from pilot to obtain the coarse data $\mathbf{\widehat{s}}_\mathrm{d}$.}
\STATE {Splice $\widehat{\mathbf{s}}_\mathrm{d}$ and $\mathbf{y}$ to real-valued using Eq.~(\ref{EQ16}).}

\STATE {Predict  $\widetilde{\mathbf{s}}_{\mathrm{FUS}}$ based on $\Theta_{\mathrm{FUS}}$ and $\widetilde{\mathbf{s}}_{\mathrm{in}}$ using Eq.~(\ref{EQ17}).}

\end{algorithmic}
\end{spacing}
\end{algorithm}
\subsection{{Estimation Feature Extraction}}
With the received signal $\mathbf{y}$, the initial features of CE are extracted by LS estimation and used as the input of the CE-Net. Using the LS estimation, the initial CFR  ${\mathbf{\widehat{h}}_{\mathrm{LS}}} \in \mathbb{C}^{N \times 1}$ is given {by}
\begin{equation}\label{EQ6}
{\mathbf{\widehat{h}}_{\mathrm{LS}}} = {\left[ {\frac{{y\left( 1 \right)}}{{{x_\mathrm{p}}\left( 1 \right)}},\frac{{y\left( 2 \right)}}{{{x_\mathrm{p}}\left( 2 \right)}}, \cdots ,\frac{{y\left( N \right)}}{{{x_\mathrm{p}}\left( N \right)}}} \right]^T},
\end{equation}
where $y(n)$ and ${x_\mathrm{p}}\left( n \right)$, $n = 1,2, \cdots ,N$, are the {received signal and transmitted pilots}, respectively. The {extracted feature i.e., $\widehat{\mathbf{h}}_{\mathrm{LS}}$, is employed} for subsequent{~enhancement of CE}.

\subsection{CE-Net based Channel Estimation}
To obtain the refined CE feature {which} is different from the conventional estimation perspective, we construct the lightweight and effective CE-Net, {which} learns the mapping relationship between input and output data. Then, a certain estimation feature, called refined {estimation feature} $\widetilde{\mathbf{h}}_{\mathrm{CE}}$, is captured through CE-Net to complement the initial estimation feature $\mathbf{\widehat{h}}_{\mathrm{LS}}$.

\subsubsection{Network Design}

According to \cite{r46}, the parameter settings of CE-Net, e.g., layer depth, layer width, and activation function, are still a challenge in the NN. Based on a large number of experimental simulations and performance tradeoffs, we determine that the CE-Net consists of $\mathcal{L}$ layers, including an input layer, two hidden layers, and an output layer. Table I summarizes the CE-Net's architecture, which is described in detail below.

 In the CE-Net, we set the neurons of the input layer and output layer as $2N$, hidden layer $1$ as $6N$, and hidden layer $2$ as $4N$ {to reduce} the complexity of CE-Net {compared to a deep network}. To avoid the overfitting problem and accelerate convergence for the CE-Net\cite{r18}, we employ the batch normalization (BN) to normalize the input layer.
   The hidden layers use rectified linear unit (ReLU) activation function, defined as ${f_a}\left( x \right) = \max \left( {0,x} \right)$, to alleviate the gradient vanishing problem\cite{r19}. And the output layer employs the linear activation function. Besides, the CE-Net refines estimation performance with these parameters.

 \begin{table*}[]

\renewcommand{\arraystretch}{1.2}
\caption{Architecture of CE-Net and FUS-Net.}
\label{table_I}
\centering
\scalebox{1.0}{
\begin{tabu}{|c|c|c|c|c|c|c|c|c|}
\tabucline[0.8pt]{-}
\multirow{2}{*}{Layer} & \multicolumn{2}{c|}{Input} & \multicolumn{2}{c|}{Hidden~1}& \multicolumn{2}{c|}{Hidden~2} & \multicolumn{2}{c|}{Output} \\ \cline{2-9}
 & \multicolumn{1}{c|}{CE-Net} & \multicolumn{1}{c|}{FUS-Net} & \multicolumn{1}{c|}{CE-Net} & \multicolumn{1}{c|}{FUS-Net} & \multicolumn{1}{c|}{CE-Net} & \multicolumn{1}{c|}{FUS-Net} & \multicolumn{1}{c|}{CE-Net} & \multicolumn{1}{c|}{FUS-Net} \\ \tabucline[0.8pt]{-}
Batch normalization &$\surd$ &$\surd$ & $\times$ & $\times$& $\times$ & - &$\times$ &$\times$ \\ \hline
Neuron number       & $2N$    & $4N$ & $6N$    & $8N$    & $4N$  &  -   & $2N$     & $2N$    \\ \hline
Activation function & - & - & ReLU & ReLU & ReLU & - & Linear & Linear  \\ \tabucline[0.8pt]{-}
\end{tabu}}
\end{table*}

 To facilitate the {real-valued CE-Net}, we reshape the complex-valued ${\mathbf{\widehat{h}}_{\mathrm{LS}}} \in \mathbb{C}^{N \times 1}$ using equation (\ref{EQ7}) to real-valued ${\mathbf{\widetilde{h}}_{\mathrm{LS}}} \in \mathbb{R}^{2N \times 1}$, which is formulated as
\begin{equation}\label{EQ7}
{\mathbf{\widetilde{h}}_{\mathrm{LS}}} = {\left[ {{\mathop{\rm Re}\nolimits} \left( {\mathbf{{\widehat{h}}}_{\mathrm{LS}}^T} \right),{\mathop{\rm Im}\nolimits} \left( {\mathbf{{\widehat{h}}}_{\mathrm{LS}}^T} \right)} \right]^T}.
\end{equation}
Next, the entries of $\mathbf{\widetilde{h}}_{\mathrm{LS}}$ form the inputs of CE-Net. Via the CE-Net, the {refined estimation feature}, denoted as ${\mathbf{\widetilde{h}}_{\mathrm{CE}}} \in \mathbb{R}^{2N \times 1}$, is given by
\begin{equation}\label{EQ8}
{\mathbf{\widetilde{h}}_{\mathrm{CE}}} = {f_{\mathrm{CE}}}\left( {{\mathbf{\widetilde{h}}_{\mathrm{LS}}},{\mathbf{\Lambda} _{\mathrm{CE}}}} \right),
\end{equation}
where ${f_{\mathrm{CE}}}\left( {\cdot} \right)$ and ${{\mathbf{\Lambda} _{\mathrm{CE}}}}$ are the CE-Net operation and its network parameters, respectively. According to (\ref{EQ8}), we refine the estimation performance without using the second-order statistics about channel.

\subsubsection{Training and Deployment}A large number of data samples are collected to train the CE-Net. Specifically, the generation of these data samples is shown below.

 For the CE-Net, the training set is presented as $\left\{ {{\mathbf{\widetilde{h}}_{\mathrm{LS}}},{\mathbf{\widetilde{h}}_{\mathrm{Label}}}} \right\}$. In this paper, the frequency-selective fading channel, i.e., $\mathbf{h}_{\mathrm{Label}}$, is derived from the widely used channel model COST2100\cite{r15}. Zadoff-Chu sequence is employed as the pilot $\mathbf{x}_\mathrm{p}$, and modulated signal $\mathbf{x}_\mathrm{d}$ is created by a quadrature-phase-shift-keying (QPSK) symbol set\cite{r21}. According to (\ref{EQ1})--(\ref{EQ4}), the set of received  signal is formed as $\left\{ \mathbf{y} \right\}$. From (\ref{EQ6}), we obtain the set $\left\{ {{\mathbf{\widehat{h}}_{\mathrm{LS}}}} \right\}$. Finally, the complex-valued sets of $\left\{ {{\mathbf{h}_{\mathrm{Label}}}} \right\}$ and $\left\{ {{\mathbf{\widehat{h}}_{\mathrm{LS}}}} \right\}$ are reshaped to the real-valued sets $\left\{ {{\mathbf{\widetilde{h}}_{\mathrm{Label}}}} \right\}$ and $\left\{ {{\mathbf{{\widetilde{h}}}_{\mathrm{LS}}}} \right\}$, respectively. We use training sets $\left\{ {{\mathbf{\widetilde{h}}_{\mathrm{LS}}},{\mathbf{\widetilde{h}}_{\mathrm{Label}}}} \right\}$ to train the CE-Net. The details are elaborated in Algorithm 1. In addition, to verify the trained network parameters during the training phase, the same generation method of training set is also used to generate a validation set\cite{r21}.

 Besides, we employ the criterion of minimizing the mean squared error (MSE) to train the CE-Net, and the loss function is expressed as
      \begin{equation}\label{EQ9}
Los{s_{{\mathrm{CE - Net}}}} = \frac{1}{{{S_1}}}\left\| {{\mathbf{\widetilde{h}}_{\mathrm{Label}}} - {\mathbf{{\widetilde{h}}}_{\mathrm{CE}}}} \right\|_2^2 + {\beta _{\mathrm{CE}}}\sum\limits_{\ell = 2}^4 {\left\| {\mathbf{W}_{\mathrm{CE}}^{\left( \ell \right)}} \right\|} _2^2,
      \end{equation}
      where $S_1$ represents the number of training samples, and ${\beta _{\mathrm{CE}}}$ denotes the regularization coefficient which is used to avoid overfitting, and $\ell$ is the layer index.

{Training set $\left\{ {{\mathbf{\widetilde{h}}_{\mathrm{LS}}}} ,\mathbf{\widetilde{h}}_{\mathrm{Label}}\right\}$  has 100,000 samples\cite{r42,r21,r52,r47}, and the batch size is set as 80 samples. Validation set of the CE-Net has 20,000 samples. The epoch number of CE-Net is set as $40$ times. Adam optimizer \cite{r22} is used as the training optimization algorithm associated with parameters $\beta_1=0.99 $ and $\beta_2=0.999 $\cite{r23}. The learning rate is set as $0.001$, and the $L_2$ regularization \cite{r24} is performed for the CE-Net.}

During training, the training operation is performed once for the CE-Net. Then, the trained network is leveraged to deploy online running.

\subsection{{Equalization Feature Extraction}}
To avoid using the second-order statistics of the noise, we employ ZF equalization to obtain the initial equalization value, which is also part of the input of FUS-Net.

 From (\ref{EQ1}), the pilot $\mathbf{x}_\mathrm{p}$ is superimposed on the modulated signal $\mathbf{x}_\mathrm{d}$. With the received signal $\mathbf{y}$, the ZF equalization is first used to highlight the initial feature for SD. Based on the refined performance of CE-Net (i.e., $\mathbf{\widehat{h}}_{\mathrm{CE}}$) and $\mathbf{y}$, the ZF equalization is formulated as
\begin{equation}\label{EQ10}
{\mathbf{\widehat{s}}_{\mathrm{ZF}}} = {\mathbf{G}_{\mathrm{ZF}}}\mathbf{y},
\end{equation}
where $\widehat{\mathbf{s}}_{\mathrm{ZF}}$ is the initial equalization feature, ${\mathbf{G}_{\mathrm{ZF}}} \in \mathbb{C}^{N \times N}$ denotes the ZF equalization matrix, which is given by
\begin{equation}\label{EQ11}
{\mathbf{G}_{\mathrm{ZF}}} = \left[ {\begin{array}{*{20}{c}}
{\frac{1}{{{{\widehat{h}}_{\mathrm{CE}}}\left( 1 \right)}}}&{}&{}&{}\\
{}&{\frac{1}{{{{\widehat{h}}_{\mathrm{CE}}}\left( 2 \right)}}}&{}&{}\\
{}&{}& \ddots &{}\\
{}&{}&{}&{\frac{1}{{{{\widehat{h}}_{\mathrm{CE}}}\left( N \right)}}}
\end{array}} \right],
\end{equation}
where ${\widehat{h}_{\mathrm{CE}}}\left( n \right)$, $n = 1,2, \cdots ,N$, is the $n$-th entry of $\mathbf{\widehat{h}}_{ \mathrm{CE}}$.

According to (\ref{EQ10}), we obtain the superimposed data and pilot $\mathbf{\widehat{s}}_{\mathrm{ZF}}$. Subsequently, we cancel the superimposed interference from pilot to obtain the coarse data $\mathbf{\widehat{s}}_\mathrm{d}$, which is given as
\begin{equation}\label{EQ12}
{\mathbf{\widehat{s}}_\mathrm{d}} = {\mathbf{\widehat{s}}_{\mathrm{ZF}}} - \sqrt {\lambda P} {\mathbf{x}_\mathrm{p}}.
\end{equation}
Then, the feature of coarse data is extracted, i.e., the coarse data $\widehat{\mathbf{s}}_\mathrm{d}$ is obtained for subsequent recovery.
\subsection{{Fusion Learning-based Symbol Detection}}
To refine the coarse data $\widehat{\mathbf{s}}_\mathrm{d}$, we draw on the idea of multimodal feature-level fusion and design a lightweight FUS-Net, which fuses coarse data feature (from the simplified equalization method using equation (\ref{EQ12})) and received signal.
\subsubsection{Network Design} After the simplified ZF equalization, the lightweight FUS-Net is utilized to refine detection performance. Similar to the CE-Net, based on extensive experiments, the FUS-Net is composed of an input layer, a hidden layer, and an output layer. The numbers of neurons in the input layer, hidden layer and output layer are $4N$, $8N$, and $2N$, respectively. The activation function is the same as the CE-Net\cite{r18}. And a BN is also used to normalize the input sets of FUS-Net, which forms the network input as zero mean and unit variance. Tabel I summarizes the FUS-Net' architecture, as described below.

The input ${\mathbf{\widetilde{s}}_{\mathrm{in}}} \in \mathbb{R}^{4N \times 1}$ of the FUS-Net is spliced by $\mathbf{{\widehat{s}}}_{\mathrm{d}}$ and $\mathbf{{y}}$, i.e.,
    \begin{equation}\label{EQ16}
\begin{array}{l}
{{\bf{\widetilde{s}}}_{{\rm{\mathrm{in}}}}} = \left[ {{\rm{Re}}\left( {{\bf{\hat s}}_{{\rm{d}}}^T} \right),{\rm{Im}}\left( {{\bf{\hat s}}_{{\rm{d}}}^T} \right),} \right.
{\left. {{\rm{Re}}\left( {{{\bf{y}}^T}} \right),{\rm{Im}}\left( {{{\bf{y}}^T}} \right)} \right]^T}.
\end{array}
    \end{equation}
Next, using the FUS-Net, the output $\mathbf{\widetilde{s}}_{\mathrm{FUS}}$ is obtained by
   \begin{equation}\label{EQ17}
{\mathbf{\widetilde{s}}_{\mathrm{FUS}}} = {f_{\mathrm{FUS}}}\left( {{\mathbf{\widetilde{s}}_{\mathrm{in}}},{\mathbf{\Lambda} _{\mathrm{FUS}}}} \right),
    \end{equation}
where ${f_{\mathrm{FUS}}}\left( {\cdot} \right)$ and ${{\mathbf{\Lambda} _{\mathrm{FUS}}}}$ are the fusion network operation and its network parameters, respectively.

\subsubsection{{{{Training and Deployment}}}}
 Similar to the CE-Net, an amount of data samples are collected to train the FUS-Net. The training details are explained as follows.

  According to (\ref{EQ16}), the input of FUS-Net $\mathbf{\widetilde{s}}_{\mathrm{in}}$ is obtained to form the real-valued fusion set $\left\{ {{\mathbf{\widetilde{s}}_{\mathrm{in}}}} \right\}$. Then, the real-valued $\left\{ {{\mathbf{\widetilde{s}}_{\mathrm{in}}}} \right\}$ and $\left\{ \mathbf{\widetilde{x}}_\mathrm{d} \right\}$ form training sets $\left\{ {{\mathbf{\widetilde{s}}_\mathrm{in}},{\mathbf{\widetilde{x}}_\mathrm{d}}} \right\}$ to train the FUS-Net. The details are described in Algorithm 1. Besides, a validation set is also needed.
   The loss function of FUS-Net is given as
      \begin{equation}\label{EQ18}
Los{s_{\mathrm{FUS - Net}}} = \frac{1}{{{S_2}}}\left\| {{\mathbf{\widetilde{x}}_\mathrm{d}} - {\mathbf{{\widetilde{s}}}_\mathrm{in}}} \right\|_2^2 + {\beta _{\mathrm{FUS}}}\sum\limits_{r = 2}^3 {\left\| {\mathbf{W}_{\mathrm{FUS}}^{\left( r \right)}} \right\|_2^2},
      \end{equation}
where $S_2$ denotes the number of training set for FUS-Net, ${\beta _{\mathrm{FUS}}}$ is the regularization coefficient, and $r$ is the number of layer.

{Training sets $\left\{ {{\mathbf{\widetilde{s}}_{\mathrm{in}}}} ,\mathbf{\widetilde{x}}_\mathrm{d}\right\}$ have 100,000 samples\cite{r42,r21,r52,r47}, and the batch sizes are set as 80. Validation sets of the FUS-Net have 20,000 samples. The epoch of FUS-Net is set as $100$. Adam optimizer \cite{r22} is used as the training optimization algorithm associated with parameters $\beta_1=0.99 $ and $\beta_2=0.999 $\cite{r23}. The learning rate is set as $0.001$, and the $L_2$ regularization \cite{r24} is used for the FUS-Net.} {For network training, we adopt mixed SNR, i.e., the training samples are generated under $0$ dB--$18$ dB.}

\subsection{Online {Deployment}}
According to the trained network parameters of CE-Net and FUS-Net through offline training, the {procedure of online running is} described in Algorithm 1. Explanations of the Algorithm 1 are given below.

In the phase of online running, the received signal $\mathbf{y}$ and the known pilot $\mathbf{x}_\mathrm{p}$ {are employed to perform the LS estimation by} using Eq.~(\ref{EQ6}). {Then}, the initial estimation $\widehat{\mathbf{h}}_{\mathrm{LS}}$ is obtained, and thus forms the network input of CE-Net (i.e., $\widetilde{\mathbf{h}}_{\mathrm{LS}}$) using Eq.~(\ref{EQ7}). With the network input $\widetilde{\mathbf{h}}_{\mathrm{LS}}$, the CE-Net refines the CE, and thus acquires the {refined estimation feature} $\widetilde{\mathbf{h}}_{\mathrm{CE}}$ with real-valued form using Eq.~(\ref{EQ8}). {The} complex-valued form of {estimation feature}, i.e., $\mathbf{\widehat{h}}_{\mathrm{CE}}$, is obtained {by extracting the real and imaginary parts from $\widetilde{\mathbf{h}}_{\mathrm{CE}}$, i.e.,}
      \begin{equation}\label{EQu1}
\left\{ \begin{array}{l}
{\mathop{\rm Re}\nolimits} \left( {{\widehat{\mathbf{h}}_{\mathrm{CE}}}} \right) = {{\mathbf{\widetilde{h}}}_{\mathrm{CE}}}\left( {1:N} \right)\\
{\mathop{\rm Im}\nolimits} \left( {{{\mathbf{\widehat{h}}}_{\mathrm{CE}}}} \right) = {\widetilde{\mathbf{h}}_{\mathrm{CE}}}\left( {N + 1:2N} \right)
\end{array} \right.,
      \end{equation}
That is, the real part and imaginary part of $\widehat{\mathbf{h}}_{\mathrm{CE}}$ are composed by extracting the first $N$ entries and the last $N$ entries of $\widetilde{\mathbf{h}}_{\mathrm{CE}}$, respectively. With the estimated $\widehat{\mathbf{h}}_{\mathrm{CE}}$, the ZF equalization is employed using Eq.~(\ref{EQ10}), and thus achieves $\widehat{\mathbf{s}}_{\mathrm{ZF}}$. Then, we cancel the superimposed interference to obtain the coarse data $\widehat{\mathbf{s}}_\mathrm{d}$ using Eq.~(\ref{EQ12}). By utilizing Eq.~(\ref{EQ16}), the real-valued $\widetilde{\mathbf{s}}_{\mathrm{in}}$ is formed based on complex-valued $\widehat{\mathbf{s}}_\mathrm{d}$ and $\mathbf{y}$. With the network input $\widetilde{\mathbf{s}}_{\mathrm{in}}$, the FUS-Net fuses the coarse data feature and received signal. Then, the FUS-Net outputs {the} detected symbol $\widetilde{\mathbf{s}}_{\mathrm{FUS}}$ {by} using Eq.~(\ref{EQ17}).

According to Algorithm 1, the refined detection $\widetilde{\mathbf{s}}_{\mathrm{FUS}}$ can be achieved from the proposed CE-Net and FUS-Net. By using the FUS-Net, the high precision detection $\mathbf{\widetilde{s}}_{\mathrm{FUS}}$ is achieved. Compared with the conventional methods, e.g., the
 MMSE CE and MMSE SD, the proposed method demonstrates a better detection performance, {e.g., a lower BER performance}. It is noteworthy that the performance of the proposed method is refined without any second-order statistic of wireless noise and channel.

\begin{table*}[]
\renewcommand{\arraystretch}{1.2}
\caption{the Analysis of Computational Complexity.}
\label{table II} \centering
\begin{tabular}{|c|c|c|}
\Xhline{0.8pt}
Method & proposed & MMSE-CE + MMSE-SD  \\
\Xhline{0.8pt}
Complexity & $84N^2$ & $6N^3+4N^2+2N$  \\
\Xhline{0.5pt}
{Case 1 ($N=32$) }& $86,016$ & $200,768$ \\
\Xhline{0.5pt}
Case 2 ($N=64$) & $344,064$ & $1,589,376$ \\
\hline
\Xhline{0.8pt}

\end{tabular}
\end{table*}

{\emph{Remark1: Battery Life and Spectral Efficiency}

{Relative to those IoT systems without employing superimposed pilot and RIS, the proposed method in this paper improves the battery life of the UE and spectral efficiency for an IoT system. Due to the superimposition mode, the energy consumption of an {IoT UE} is significantly reduced given the same transmitted power. Besides, compared with the {IoT systems} without RIS, the application of RIS increases the communication reliability and thus improves the energy consumption as well for the similar communication quality. For an IoT system with limited bandwidth, the approach of using superimposed pilots in this paper effectively improves spectral efficiency. Therefore, compared with those IoT systems without employing superimposed pilot and RIS, the proposed superimposed pilot-based CE {with RIS assistance} significantly prolongs {UE's battery life} {and improves the {spectral efficiency of} IoT systems.}}

The proposed method adopts superimposed pilot mode, and the UE does not need extra resources for pilot transmission. Thus, compared with CE methods of non-superimposed pilot \cite{r39}, \cite{r38}, the spectral efficiency is improved. Meanwhile, the energy consumption at the UE is reduced due to the fact that extra energy for pilot transmission is avoided. Table III shows the comparison of bandwidth resource occupation and energy consumption between the non-superimposed
pilot-based CE method \cite{r39}, \cite{r38} and the proposed method in this paper.


By denoting the energy consumption of the non-superimposed pilot-based CE as $E_{\mathrm{NonSup}}$, then we have
\begin{equation}\label{EQ999}
{E_{{{\mathrm{NonSup}}}}} = \left( {{N_{{{\mathrm{data}}}}} + {N_{{{\mathrm{Pilot}}}}}} \right){T_0}{P},
\end{equation}
{where $N_{\mathrm{data}}$ denotes the number of data symbols, $N_{\mathrm{Pilot}}$ represents the number of pilot symbols, $T_0$ is the symbol duration, and $P$ stands for the transmitted power.}

Compared with {the non-superimposed pilot-based CE \cite{r39}, \cite{r38}, the superimposed pilot mode saves the energy consumption of UE due to the fact that the extra energy consumption for pilot transmission is avoided. In this paper, the energy consumption of the proposed scheme is denoted as $E_{\mathrm{Prop}}$, which can be expressed as}
\begin{equation}\label{EQ99}
{E_{{\rm{Prop}}}} = {N_{{\rm{data}}}}{T_0}\left( { {\left( {1{\rm{ - }}\lambda } \right)P} } \right) + {N_{\mathrm{Pilot}}}{T_0}\left( { {\lambda P} } \right),
\end{equation}
where $\lambda$ denotes the power proportional coefficient. Then, compared with the non-superimposed pilot-based CE, the saved energy consumption by using the the proposed method is given by
\begin{equation}\label{EQ990}
{E_{{\rm{NonSup}}}} - {E_{{\rm{Prop}}}}  = {N_{{\rm{data}}}}{T_0}\left( { {\lambda P} } \right) + {N_{\mathrm{Pilot}}}{T_0}\left( { {\left( {1{\rm{ - }}\lambda } \right)P} } \right).
\end{equation}
In terms of bandwidth resource occupation, the proposed method transmits pilot in a superimposed {manner}, {in which the time of bandwidth occupation is ${N_{{\rm{data}}}}{T_0}$. {In contrast, the bandwidth resource occupation of non-superimposed pilot-based CE} is $\left( {{N_{{\rm{data}}}} + {N_{{\rm{Pilot}}}}} \right){T_0}$.}
Thus, relative to non-superimposed pilot-based CE, the proposed method reduces the bandwidth resource occupation, which can be given by $\left( {{N_{{\rm{data}}}} + {N_{{\rm{Pilot}}}}} \right){T_0} - {N_{{\rm{data}}}}{T_0} = {N_{{\rm{Pilot}}}}{T_0}$.
By considering {the case} where ${N_{{\rm{data}}}} = 32$ and ${N_{{\rm{Pilot}}}} = 32$, it can be seen from Table III that compared with the CE method based on non-superimposed pilot, the proposed method reduces the bandwidth resource occupation and energy consumption. To sum up, compared with non-superimposed pilot-based CE, the proposed method improves the spectral and energy efficiency of RIS-assisted IoT systems.
\begin{table*}[]
\renewcommand{\arraystretch}{1.2}
\caption{Bandwidth resource occupation and energy consumption.}
\label{table_V}
\centering
\scalebox{1.0}{
\begin{tabu}{|c|c|c|c|c|}
\tabucline[0.8pt]{-}
\multirow{2}{*}{Method} & \multicolumn{2}{c|}{Bandwidth resource usage} & \multicolumn{2}{c|}{Total energy consumption} \\ \cline{2-5}
 & \multicolumn{1}{c|}{Expression} & \multicolumn{1}{c|}{Example} &  \multicolumn{1}{c|}{Expression} & \multicolumn{1}{c|}{Example}\\ \tabucline[0.8pt]{-}
 Non-superimposed pilot&$\left( {{N_{{\rm{data}}}} + {N_{{\rm{Pilot}}}}} \right){T_0}$ &$64{T_0}$ & ${{{{{}}}}}  \left( {{N_{{{\mathrm{data}}}}} + {N_{{{\mathrm{Pilot}}}}}} \right){T_0}{P}$ & $64{T_0}{P_0}$ \\ \hline
       Superimposed pilot& $\left( {{N_{{\rm{data}}}}} \right){T_0}$    & $32{T_0}$ & ${{{{}}}} {N_{{\rm{data}}}}{T_0}\left( { {\left( {1{\rm{ - }}\lambda } \right)P} } \right) + {N_{\mathrm{Pilot}}}{T_0}\left( { {\lambda P} } \right)$    & $32{T_0}{P_0}$       \\ \tabucline[0.8pt]{-}
\end{tabu}}
\end{table*}

{In addition to the benefits of \emph{Remark 1}, the {proposed superimposed pilot-based CE with RIS assistance in this {paper} {also reduces} the computational complexity and {processing delay} at the BS, compared with the IoT systems without employing superimposed pilot and RIS}. The specific analysis and comparison of computational complexity and processing delay at the BS are presented in Section IV.}

\section{{Complexity and Running Time Analyses}}

For convenience, the simplified expression {is as follows.}
\begin{itemize}
  \item ``LS-CE'', ``MMSE-CE'' and ``CE-Net'' {are used to represent the} ``LS channel estimation'', ``MMSE channel estimation'', and ``proposed CE-Net'', respectively.
  \item ``MMSE-CE + MMSE-SD'', ``CE-Net + ZF'' and ``proposed'' {are utilized to stand} the ``MMSE channel estimation followed by MMSE equalization'', ``proposed CE-Net followed by ZF equalization'', and ``proposed CE-Net followed by FUS-Net'', respectively.
\end{itemize}

\subsubsection{Computational Complexity}
{As the most common criterion, the computational complexity of NN is described in terms of weight number and floating-point operations (FLOPs) \cite{r47}. In this paper, we employ these criteria to compare the computational complexity, which is elaborated in Table II and some details are given as follows. }

{According to the computing method given in} \cite{r47}, the total NN weight number of the proposed CE-Net and FUS-Net is $28N^2+8N$, and the total FLOPs number is $56N^2-8N$. Thus, the {computational complexity} of the proposed method (including the CE-Net and FUS-Net) is $28N^2+8N + 56N^2-8N = 84N^2$. As shown in Table II, the proposed method has lower computational complexity than that of the ``MMSE-CE + MMSE-SD''. {For the case where $N = 32$, i.e., case 1 in Table II, the computational complexity of the ``MMSE-CE + MMSE-SD'' is 200,768, whereas the computational complexity of the ``proposed'' is 86,016. When $N = 64$ (i.e., case 2 in Table II), the computational complexity of the ``MMSE-CE + MMSE-SD'' is 1,589,376, while the computational complexity of the ``proposed'' is 344,064. On the whole, compared with the ``MMSE-CE + MMSE-SD'', the proposed method reduces {the computational complexity and thus obtains the corresponding improvement for energy-consuming.}
\begin{figure}[H]
\centering
\includegraphics[scale=0.6]{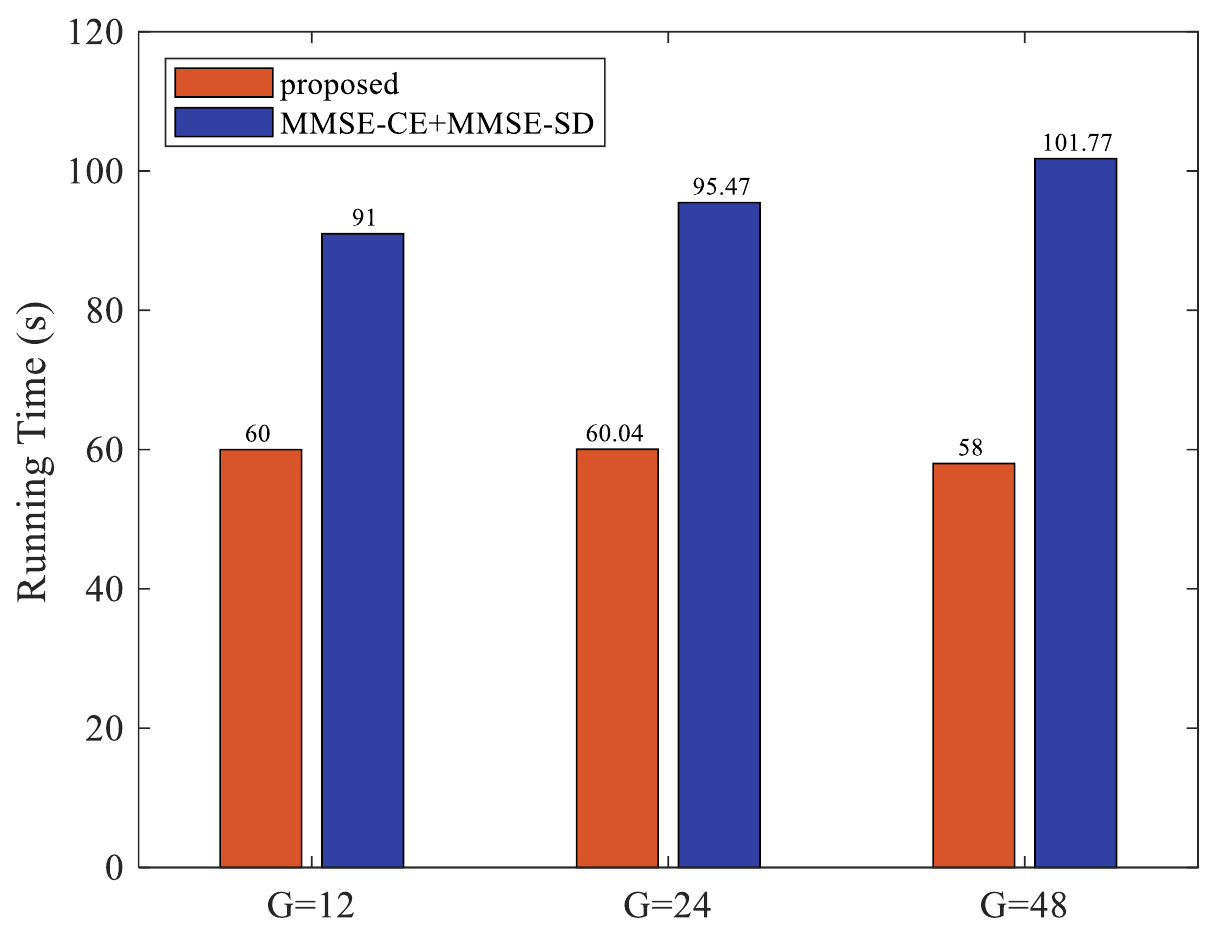}
\caption{Running time comparison between ``proposed'', and ``MMSE-CE+MMSE-SD'' for $3\times10^4$ experiments, where $G=12$, $G=24$, and $G=48$ are discussed.}
\label{fig9}
\end{figure}
\subsubsection{{Running Time}}
The training of the proposed
method is obtained on a server with Intel Xeon(R) E5-2620 CPU 2.1GHz$\times$16, and the results are got by using
MATLAB simulation on the server CPU due to the lack of a
GPU solution {for the ``MMSE-CE + MMSE-SD''}. The details of running time are discussed {in Fig.~\ref{fig9}. For} the case that $G=12$, {the total online running time of the ``proposed'' is about $60$ seconds for the two networks,} whereas that of the ``MMSE-CE + MMSE-SD'' is about {$91$ seconds}. It can be seen that the online running time of the proposed method is less than {that of the ``MMSE-CE + MMSE-SD'', prolonging the battery life of the UE as well}.

{Thus, compared with the ``MMSE-CE + MMSE-SD'', the ``proposed'' significantly reduces their computational complexity and running time.}

\begin{figure}[H]
\centering
\includegraphics[scale=0.6]{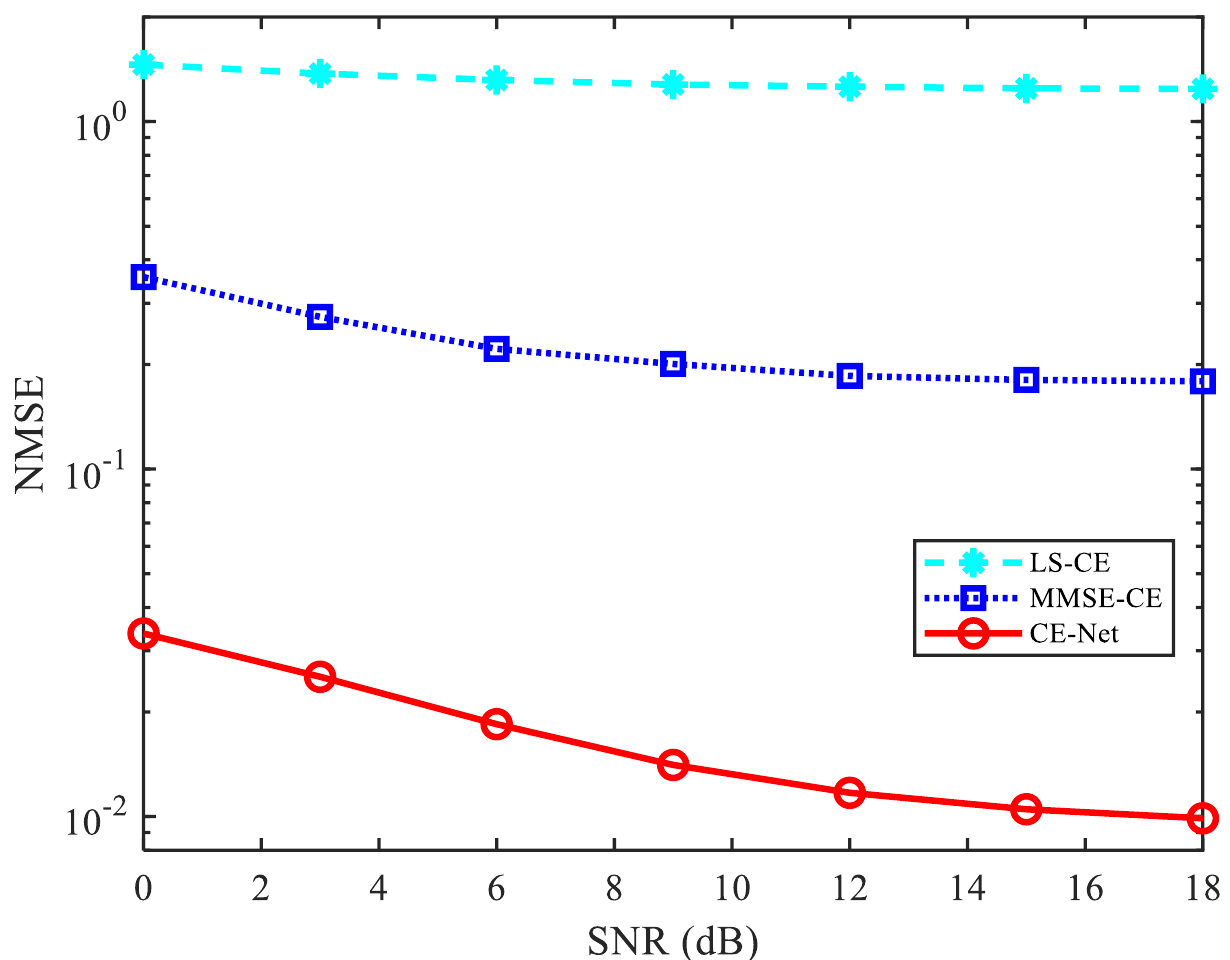}
\caption{ NMSE comparison of ``LS-CE'', ``MMSE-CE'', ``CE-Net'', where SNR varies from $0$ to $18$ dB . }
\label{figNMSE}
\end{figure}
\begin{figure}[H]
\centering
\includegraphics[scale=0.6]{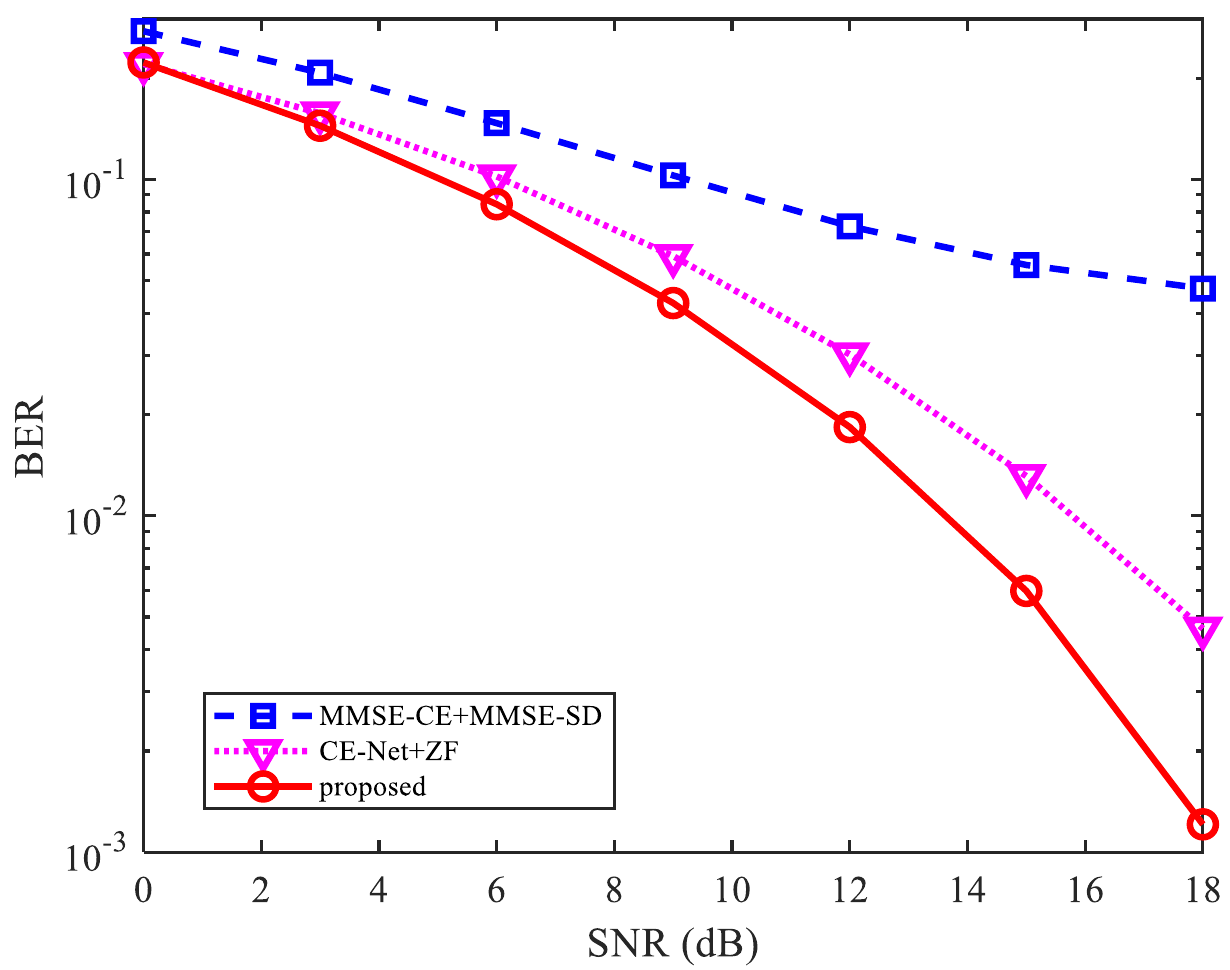}
\caption{ BER comparison of ``proposed'', ``MMSE-CE+MMSE-SD'', ``CE-Net+ZF'', where SNR varies from $0$ to $18$ dB. }
\label{figBER}
\end{figure}

\section{Simulation Results and Analysis}

In this Section, numerical results of the proposed method are given. The basic parameters and definitions involved in the simulations are presented in Section V-{A}. Then, in Section V-{B}, the simulation results verify the effectiveness of the proposed method. Finally, the parameters robustness analyses are elaborated in Section V-{C}.

\subsection{Parameters and Definitions }
In all the experiments, unless otherwise specified, the following basic parameters are used. The pilot is Zadoff-Chu sequence\cite{r14}, $L=5$, $N=32$, $\lambda=0.15 $, and $G=12$. The channel is generated by channel model COST2100\cite{r15}, and the transmitted data symbol is modulated by QPSK\footnote{The modulation with a higher modulation order is also suitable for the proposed method.} modulation.
 The signal to noise ratio (SNR) in decibel (dB) is expressed as\cite{c21}
\begin{equation}\label{19}
{\mathrm{SNR} = 10{\log _{10}}\left( {\frac{{{P{}}}}{{\sigma _w^2}}} \right)},
\end{equation}
where ${P_{{}}}$ is the total transmitted power of superimposed data and pilot, which is equal to the sum of data power ${P_\mathrm{d}}$ and pilot power ${P_\mathrm{p}}$. In these simulations, ${P_\mathrm{d}} = 0.85{P_{{}}}$ and ${P_\mathrm{p}} = 0.15{P_{{}}}$.

The NMSE is utilized to evaluate the CE performance, which is defined as\cite{c21}
\begin{equation}\label{100}
\mathrm{NMSE} = \frac{{\left\| {{\mathbf{{\hat h}}_{\mathrm{CE}}} - \mathbf{h}} \right\|_2^2}}{{\left\| \mathbf{h} \right\|_2^2}}.
\end{equation}

\begin{figure*}[t]
\centering
\includegraphics[scale=0.45]{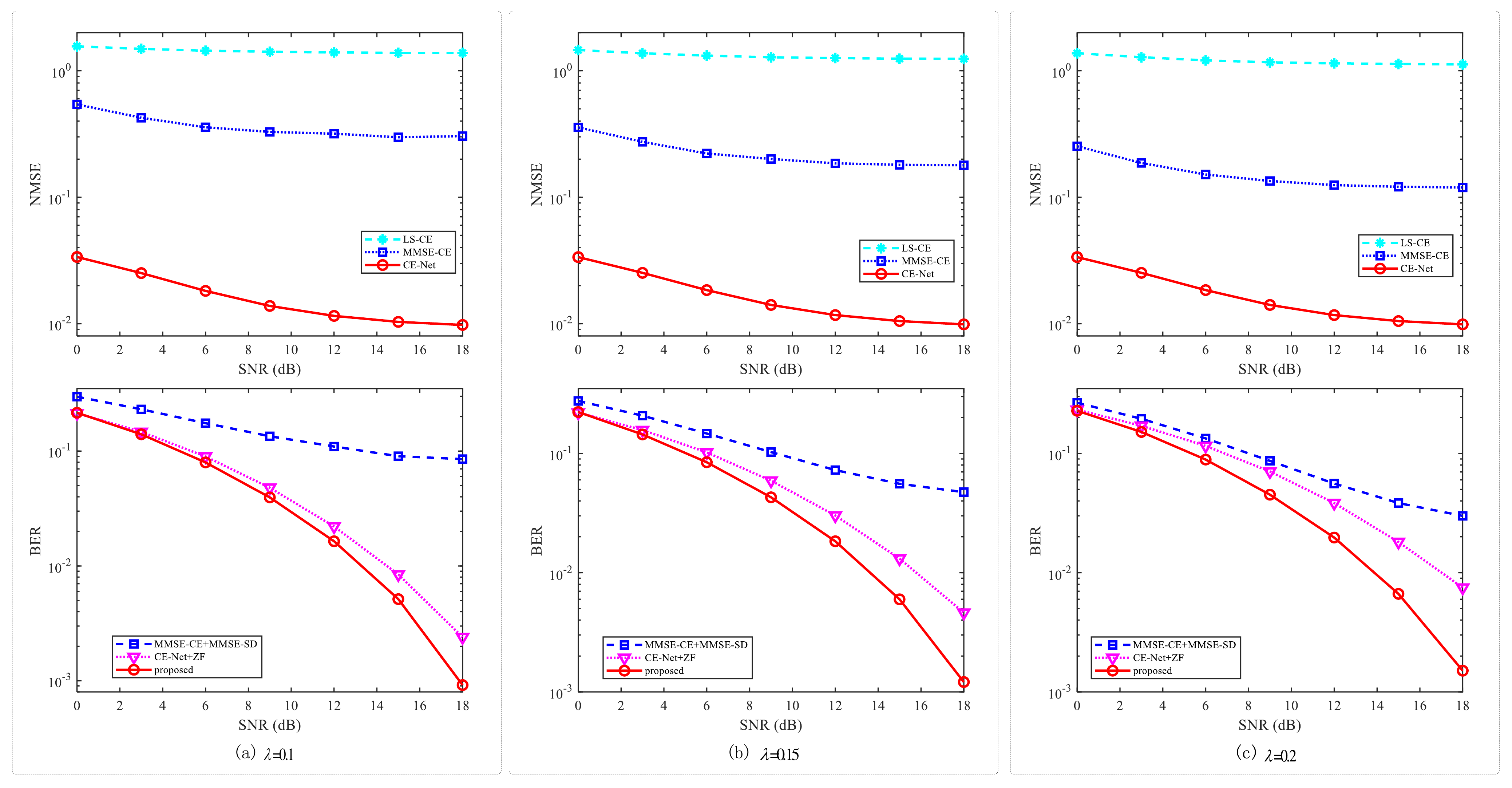}
\caption{ NMSE and BER performance against the impact of $\lambda$, where $\lambda=0.1$, $\lambda=0.15$, and $\lambda=0.2$ are considered, respectively. }
\label{fig3}
\end{figure*}

\subsection{NMSE Performance Analysis}
We {validate} the effectiveness of the {proposed CE-Net} in terms of the NMSE curves in Fig.~\ref{figNMSE}. As shown in Fig.~\ref{figNMSE}, {the values of NMSE of} ``LS-CE'' and ``MMSE-CE'' are much higher than that of the ``CE-Net'' {for all given SNRs}. For example, the NMSE of the ``CE-Net'' is less than $1\times10^{-2}$ for the case of {SNR~}$ = 18$ {dB}, while the NMSE of the ``MMSE-CE'' is $2\times10^{-1}$ and ``LS-CE'' is higher than $1\times10^{-0}$ at the same SNR. The reason of the poor performance of the ``LS-CE'' is that the LS estimation is sensitive to the {noise and interference}. The superimposed pilot is equivalent to introducing the {superimposed interference}, which results in an unsatisfactory LS estimation. {The NMSE of ``MMSE-CE'' is lower than that of the ``LS-CE'' due to the utilization of the second-order statistical information about the channel and noise, which is at the cost of higher computational complexity. However, the NMSE} of ``MMSE-CE'' is still {unsatisfying} due to the influence of superimposed interference. {In contrast, the developed CE-Net effectively alleviates} the impact of superimposed interference by exploiting its learning ability of de-noising (suppressing the superimposed interference and noise) and feature extraction (learning the feature of wireless channels). {Thus,} compared with the linear solution estimated by {LS and MMSE-based CE}, the developed CE-Net learns a nonlinear solution orienting the LS solution, which improves the CE's {NMSE performance.}

\subsection{BER Performance Analysis}
Since the pilot $\mathbf{x}_\mathrm{p}$ is superimposed on the modulated symbol $\mathbf{x}_\mathrm{d}$, it is necessary to verify whether the superimposed interference (from the pilot) degrades the detection performance of the data symbol. In this paper, the BER is used as the metric of the detection performance and plotted in Fig.~\ref{figBER}.
We utilize ``MMSE-CE+MMSE-SD'' and ``CE-Net+ZF'' as the baseline to evaluate the effectiveness of BER for the ``proposed'' method. As shown in Fig.~\ref{figBER}, the {BER curve of} the ``proposed'' is {much smaller} than that of the ``MMSE-CE+MMSE-SD''. For example, {for the case where {SNR~}$=14${dB}}, the BER of the ``proposed'' is less than $1\times10^{-2}$ while the BER of the ``MMSE-CE+MMSE-SD'' {is about} $6.5\times10^{-2}$. Furthermore, the BER of the ``CE-Net+ZF'' is smaller than that of the ``MMSE-CE+MMSE-SD''. {One of the main reasons is that the poor NMSE performance of the ``MMSE-CE+MMSE-SD'' affects the subsequent detection performance. The error of CE is propagated to the detection stage and thus degrades the detection performance of the ``MMSE-CE+MMSE-SD''.}

 With the {superior} learning ability of the CE-Net, the {NMSE performance of the ``proposed'' is improved by the CE-Net}, {to improve its BER performance}. At the same time, we can see that the ``proposed'' achieves a {smaller value of BER}  {than} the ``CE-Net+ZF''. For example, when SNR {$=18$ {dB}}, the BER of the ``proposed'' is $1.2\times10^{-3}$ while the BER of the ``CE-Net+ZF'' {reaches} $5\times10^{-3}$.                                                                                                                                                                                                                                                                                                                                                                                                                                                                                                                                                                                                                                                                                                                              Because there is an additional data fusion network {FUS-Net} {in the ``proposed'',} it is more powerful to capture additional features {for SD and thus effectively improve its BER performance}.

\subsection{Robustness Analysis}
In this subsection, the robustness of the ``proposed'' method is analysed for {the impacts of varying parameters}, i.e., the power proportional coefficient $\lambda $, and the number of multi-path $L$. For the convenience of analysis, only one impact parameter is changed, and other basic parameters remain unchanged as given in Section V-A.

\text{\\ \\ \\}
\begin{figure*}[th]
\centering
\includegraphics[scale=0.45]{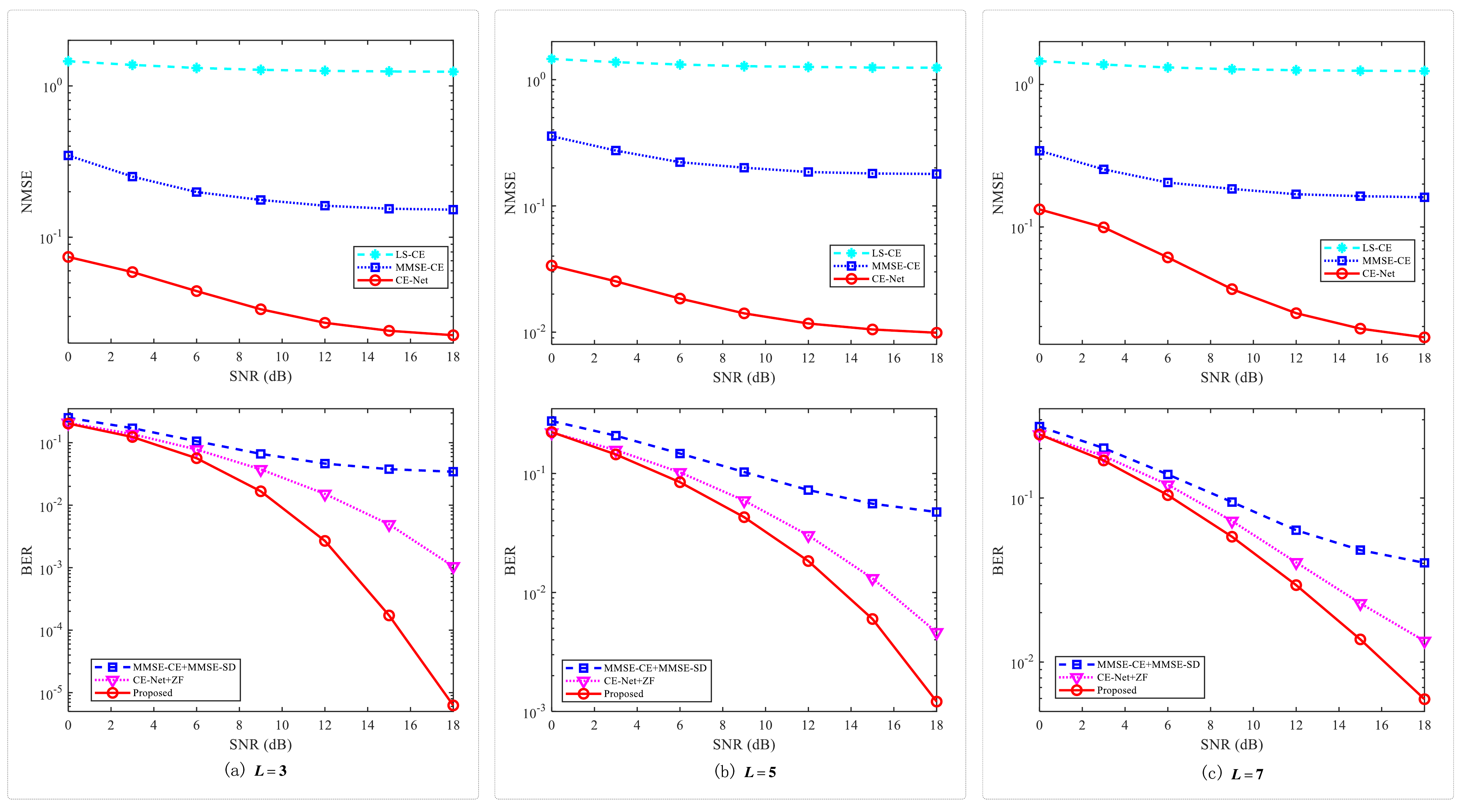}
\caption{ NMSE and BER performance against the impact of $L$, where $L=3$, $L=5$, and $L=7$ are considered, respectively. }
\label{fig4}
\end{figure*}

\subsubsection{Robustness Against $\lambda$}In general, different power proportional coefficient $\lambda$ will result in different performance {of CE and SD} for the superimposed signals. In order to demonstrate the robustness of the ``proposed'' method against $\lambda$, the NMSE of CE and the BER of SD are described in Fig.~\ref{fig3}.

{From Fig.~\ref{fig3}}, as the increase of $\lambda$ (increase from $0.1$ to $0.2$), the CE's NMSEs of ``LS-CE'' and ``MMSE-CE'' decrease. Although the decline of NMSE is not obvious, the decreasing trend is still observed. For example, when SNR {$=12${dB}} and $\lambda$ {changes} from $0.1$ to $0.2$, the ``MMSE-CE'' {changes} from $3\times10^{-1}$ to $1.2\times10^{-1}$. The likely reason is that the CE performance  is improved due to the increased pilot power.
 Meanwhile the NMSE performance of the ``proposed'' remains stable and is {smaller} than  ``LS-CE'' and ``MMSE-CE'' with the {increase} of $\lambda$. For example, for the case {where} SNR {$=12${dB}} and $\lambda=0.15$, {the values of NMSE are higher than $1\times10^{0}$ and $2\times10^{-1}$ for ``LS-CE'' and ``MMSE-CE'', respectively. By contrast,}~the NMSE of {the} ``proposed'' is about $1\times10^{-2}$.

{ With the increase of $\lambda$, the BER performance of the ``CE-Net+ZF'' and ``proposed'' deteriorate slightly. For example, for the case where SNR $=18$ dB and $\lambda=0.1$, the values of BER are about $2.2\times10^{-3}$ and $1\times10^{-3}$, respectively. While for the case where SNR $=18$ dB and $\lambda=0.2$, the values of BER are about $8\times10^{-3}$ and $2\times10^{-3}$, respectively. However, the BER of the ``proposed'' remains much smaller than those of ``CE-Net+ZF'' and ``MMSE-CE+MMSE-SD'' for each given SNR and $\lambda$. Thus, against the impact of $\lambda$, the ``proposed'' improves the BER performance when compared with the ``CE-Net+ZF'' and ``MMSE-CE+MMSE-SD''. }

  {On the whole, when compared with the ``CE-Net+ZF'' and ``MMSE-CE+MMSE-SD'', the ``proposed'' enhances the NMSE and BER performance against the variation of $\lambda$.}

\subsubsection{Robustness Against $L$}The performance is usually influenced by the number of multi-path, i.e., $L$. To illuminate the robustness against the impact of {$L$,} the {performance comparison} {is shown in Fig.~\ref{fig4}}, where $L=3$, $L=5$, and $L=7$ are considered.
As shown in Fig.~\ref{fig4}, the varying of NMSE is not regular with the enlargement of $L$. The reason is that the performance of NMSE is not so directly related to {the values of $L$}. {Even so}, we can see that no matter how the {values of $L$} change, using the ``CE-Net'' {achieves} the {minimum value of NMSE}, {presenting the best NMSE performance}. For example, when SNR {$=12$dB }and $L=5$, {the NMSE values of ``LS-CE'' and ``MMSE-CE'' are respectively higher than $1\times10^{0}$ and $2\times10^{-1}$,} while the NMSE of ``CE-Net'' is about $1.2\times10^{-2}$. This reflects that the CE-Net improves the {NMSE performance} compared with the conventional methods {of ``LS-CE'' and ``MMSE-CE'' against the variations of $L$}.

Besides, from Fig.~\ref{fig4}, {compared with the ``MMSE-CE+MMSE-SD'' and ``CE-Net+ZF'',} the ``proposed'' achieves {smaller} BER for each given $L$. For example, for the cases {of} SNR {$=18$} dB and $L=5$, {BERs of the ``MMSE-CE+MMSE-SD'' and ``CE-Net+ZF'' are {about} $5.5\times10^{-2}$ and $4.5\times10^{-3}$} respectively, while the BER of ``proposed'' is {smaller than} $2\times10^{-3}$. This reflects that {the ``proposed''} improves the BER compared with the {``MMSE-CE+MMSE-SD'' and ``CE-Net+ZF'' against} the variation of $L$. Besides, it is worth noting that for the case of $L=5$, each of the CE methods achieves the smallest NMSE, yet they cannot achieve the best detection performance. This is because in the case of superimposed pilots, although the estimation performance is improved, the detection performance is not necessarily improved proportionally due to the influence of superimposed interference. {Thus, an effective option is to make a tradeoff between NMSE performance and BER performance for the superimposed pilot-based method}.

 {Therefore,} against the impact of $L$, Fig.~\ref{fig4} shows that both of the NMSE and BER performance are improved by ``proposed'' when compared with the ``CE-Net+ZF'' and ``MMSE-CE+MMSE-SD''.}

\section{Conclusion}
In this paper, a superimposed pilot-based CE with RIS-assisted mode is proposed for IoT systems. The spectral efficiency and the energy consumption are improved by employing the superimposed pilot, and the issue of blocked propagation paths is alleviated by deploying RIS. Besides, non-NN and NN-based modes are integrated at the BS to form lightweight networks, which effectively reduce the computational complexity and processing delay. Compared with the conventional methods, the proposed solution shows its effectiveness and robustness in improving the NMSE and BER performance. In our future works, we will consider the influence of RIS materials on CE.


%
\bibliographystyle{ieeetran}


\end{document}